\documentclass[12pt]{article}
\begin{document}
\title{{\bf ANTHROPIC ESTIMATES
\\ OF THE CHARGE AND MASS OF THE PROTON}
\thanks{Alberta-Thy-03-04, hep-th/0203051}}
\author{
Don N. Page
\thanks{Internet address:
don@phys.ualberta.ca}
\\
Institute for Theoretical Physics\\
Department of Physics, University of Alberta\\
Edmonton, Alberta, Canada T6G 2J1
}
\date{(2003 Feb. 13)}
\maketitle
\large

\begin{abstract}
\baselineskip 18 pt

	By combining a renormalization group argument
relating the charge $e$ and mass $m_p$ of the proton
by $e^2 \ln{m_p} \approx -\,0.1\pi$ (in Planck units)
with the Carter-Carr-Rees anthropic argument that gives
an independent approximate relation $m_p \sim e^{20}$
between these two constants, both can be crudely estimated.
These equations have the factor of $0.1\pi$ and the exponent of 20
which depend upon known discrete parameters
(e.g., the number of generations of quarks and leptons,
and the number of spatial dimensions),
but they contain {\it no} continuous observed parameters.
Their solution gives the charge of the proton correct to within about 8\%,
though the mass estimate is off by a factor of about 1000
(16\% error on a logarithmic scale).
When one adds a fudge factor of 10 previously given by Carr and Rees,
the agreement for the charge is within about 2\%,
and the mass is off by a factor of about 3
(2.4\% error on a logarithmic scale).
If this 10 were replaced by 15, the charge agrees
within 1.1\% and the mass itself agrees within 0.7\%.

\end{abstract}
\normalsize
\baselineskip 18 pt
\newpage

\section{Introduction}

	One prominent idea of recent decades is that
some of the `constants of physics' may only be constant
(or nearly so) in our observable region of the universe
(our `subuniverse')
but may take on a range of values in the entire universe
(or `multiverse,' though I shall eschew this terminology here).
Two examples to be considered here are the electromagnetic
fine structure constant, $\alpha = e^2/(4\pi\epsilon_0\hbar c)$
(the square of the charge of the proton, $e$, in
the Planck units $\hbar = c = G = 4\pi\epsilon_0 = 1$
I shall use here), 
and the gravitational fine structure constant for the proton,
$\alpha_G = Gm_p^2/(\hbar c) \equiv (m_p/m_P)^2$
(the square of the mass of the proton, $m_p$, in Planck units).
If these `constants' vary widely over the entire universe,
most of the universe may be unsuitable for observers,
and hence one might expect that observed values would
statistically have a narrower range.
One goal of a complete theory of physics and of observership
would be to calculate the statistical distribution of
the observed values of such `constants.'

	Of course, we are nowhere near achieving such a goal.
However, here I wish to use come crude hypotheses and arguments
about the conditions that might be conducive for typical observers
in order to come up with some rough values that typical observers
might be expected to see for the electromagnetic and gravitational
fine structure constants.  (Such arguments often go under the name
of anthropic arguments, but they are really about observership
rather than about {\it anthropos}.  Because these arguments can be made
only very crudely at present, they are often deprecated,
but I shall take the view that they are just the beginning hints
of what should in principle be possible with a complete theory
of the universe, including the physical conditions for observership.)

	Rather remarkably, one finds that
without using any observational data except for certain discrete
integers (e.g., the dimension of space, the number of generations
of quarks and leptons, etc.), one can get crude estimates
for typical observed values of both of these parameters
that are in the right ballpark to explain our particular observations.
In this paper I shall show how one can get a crude estimate
for the magnitude of the charge of the proton
($e \equiv \sqrt{\alpha}$ in Planck units)
that differs from what we observe by only a few percent.
One can also get a somewhat more crude estimate
for the mass of the proton ($m_p \equiv \sqrt{\alpha_G}$ in Planck units)
that differs by about 3 orders of magnitude
from the value that we observe.
However, since the observed value is about 19 orders
of magnitude smaller than unity,
on a logarithmic scale the crude estimate is not that far off.

	Furthermore, one just needs a single previously-published fudge
factor of 10 to reduce the error to about 2\% for the charge of the proton
(or to within 1.1\% error for the charge
and to within 0.7\% error for the mass if this 10 were replaced by 15).
Of course, it is a cheat
to take these revised values as predictions,
but to the degree that it is not implausible that a factor
of the order of 10 or 15 could arise from a more detailed calculation,
it is not implausible that such a calculation could give
a statistical distribution of observed values of the
charge and mass of the proton
such that our observations of these values would be typical.
In other words, the `errors' of the crude estimates
previously mentioned,
which do not use as input any observed continuous parameters
like the fudge factors of the present paragraph,
seem small enough not to be fatal to the form of the argument. 

\section{The Carter-Carr-Rees anthropic argument}

	We will need two independent formulas relating
the charge and mass of the proton in order to deduce
approximations for their values.  First I shall give
the Carter-Carr-Rees formula,
and then in the next section I shall add my own renormalization
group argument to get a new formula that one can combine
with the Carter-Carr-Rees one to get actual estimates
for the charge and mass of the proton.

	Brandon Carter \cite{Carter} noted that typical observers
may live on planets, as we do, and that planetary formation
may depend upon the formation of convective red stars.
For the existence of these he deduced that
 \begin{equation}
 \alpha_G \equiv m_p^2 \, \stackrel{<}{\sim} \,
  \alpha^{12}\left({m_e\over m_p}\right)^4,
 \label{eq:1}
 \end{equation}
where $m_e$ is the mass of the electron.
Then Bernard Carr and Martin Rees \cite{Carr-Rees}
noted that if $\alpha_G$ were a lot smaller than the
right hand side of Eq. (\ref{eq:1}),
one might not get supernovae that are necessary
for producing the elements needed for observers like us.
Putting these two arguments together,
one may conclude that for observers like us
that live on planets and have a complex chemistry,
one may need
 \begin{equation}
 \alpha_G \sim \alpha^{12}\left({m_e\over m_p}\right)^4.
 \label{eq:2}
 \end{equation}
 
 	One might alternatively write this relation as
 \begin{equation}
 \alpha_G = C\alpha^{12}\left({m_e\over m_p}\right)^4
 \label{eq:3}
 \end{equation}
with a `Carter constant' $C$ that is predicted to be
observed to be of the order of unity by typical observers.
We observe it to be $C \approx 2.944$ in our sub-universe \cite{PDG}.
 
	In their Eq. (58), Carr and Rees \cite{Carr-Rees}
also cited Carter \cite{Carter2} in noting
four key coincidences of nuclear physics that may be
necessary for observers like us.  Together they imply
what Carr and Rees write, in Eq. (45) of their paper \cite{Carr-Rees},
as
 \begin{equation}
 {m_e \over m_p} \sim 10 \, \alpha^2.
 \label{eq:4}
 \end{equation}
The 10 is really just an order-of-magnitude estimate
for a numerical coefficient that is not predicted precisely,
a `fudge factor' as it were, so to get relations
without including unpredicted factors, one might write this
relation merely as
 \begin{equation}
 {m_e \over m_p} \sim \alpha^2.
 \label{eq:5}
 \end{equation}
Alternatively, one might write this relation as
 \begin{equation}
 {m_e \over m_p} = R \, \alpha^2
 \label{eq:6}
 \end{equation}
with a `Carr-Rees constant' $R$ that is predicted
to be somewhat larger than unity.
To give an order of magnitude estimate, Carr and Rees say $R \sim 10$.
We actually observe it to be $R \approx 10.227290$ in our subuniverse
\cite{PDG}.

	If we now combine these two relations
without the numerical factors of $C$ and $R$, we get
what I shall call the basic Carter-Carr-Rees relation
 \begin{equation}
 \alpha_G \sim \alpha^{20},
 \label{eq:7}
 \end{equation}
which is also listed explicitly as Eq. (57) of \cite{Carr-Rees}.

	If one does include the fudge factor of 10
that Carr and Rees include in their Eq. (45),
copied as Eq. (\ref{eq:4}) above,
then one gets what I shall call the fudged
Carter-Carr-Rees relation
 \begin{equation}
 \alpha_G \sim 10^4\,\alpha^{20}.
 \label{eq:8}
 \end{equation}
Alternatively, one can write
 \begin{equation}
 \alpha_G = E \,\alpha^{20} = (F \alpha)^{20},
 \label{eq:9}
 \end{equation}
where $E = F^{20} = C R^4$.
In our subuniverse, we observe $E \approx 32205$,
or $F \approx 1.6803 \approx 2^{0.7487}$ \cite{PDG}.
Since $F$ is fairly close to $2^{3/4}$,
as a mnemonic device one can remember
a fairly good approximation to Eq. (\ref{eq:9})
in our subuniverse as the ``2-3-4'' formula
 \begin{equation}
 \alpha_G \approx (2^{3/4} \alpha)^{20}.
 \label{eq:9b}
 \end{equation}

	Thus we have one rough relation
between $\alpha \equiv e^2$ and $\alpha_G \equiv m_p^2$,
or between the charge $e$ and mass $m_p$ of the proton.
To get an estimate for both values, we need another relation.
This I shall provide by a crude renormalization-group (RG)
analysis of supersymmetric (SUSY)
Grand Unified Theory (GUT) couplings
in the Minimal Supersymmetric Standard Model (MSSM).

	But before I do that, I should note that the Carr-Rees
paper, though preceding most of the development of the MSSM,
did also propose a second relation between $\alpha$
and $\alpha_G$ based upon some arguments about
a self-consistent quantum electrodynamics, namely Eq. (54)
of \cite{Carr-Rees}, which I shall call the Carr-Rees log relation:
 \begin{equation}
 \alpha^{-1} \sim \log \, \alpha_G^{-1}.
 \label{eq:10}
 \end{equation}

	Although Carr and Rees did not carry this further
in their paper, it is trivial to combine Eqs. (\ref{eq:7})
and (\ref{eq:10}) [their Eqs. (57) and (54) respectively]
to estimate that $\alpha^{-1} \sim 90$, which is about 66\%
of the value we observe in our subuniverse \cite{PDG},
$\alpha_{\mathrm{exp}}^{-1} \approx 137.036000$,
and to estimate that $\alpha_G^{-1} \sim 1.2 \times 10^{39}$,
about 7 times the value we observe in our subuniverse \cite{PDG},
$\alpha_{G\ \mathrm{exp}}^{-1} \approx 1.693 \times 10^{38}
 \approx 2^{126.993} \approx 2^{2^{2^{2^2-1}-1}-1}-1$
(the 12th smallest Mersenne prime and largest prime found
without computers, in 1876 by Edouard Lucas \cite{Lucas}).
It is remarkable that even these very crude estimates
give values for the charge and mass of the proton that
are within the right ballpark for our observations.

\section{The renormalization group argument}

	Now I shall show how to deduce a better relation
between $\alpha^{-1}$ and $\ln \alpha_G^{-1}$
by using a renormalization group analysis of
supersymmetric grand unified theories, e.g.,
that of William Marciano and Goran Senjanovi\'c \cite{MarCen}.

	I shall assume that there are $n_g = 3$ generations
of quarks and leptons and $N_H = 2$ relatively light Higgs doublets
in low-energy $SU(3)\times SU(2)\times U(1)$,
which are discrete choices that themselves might by justified
in the future by some anthropic argument that typical observers
would see those values.

	However, I shall not use any observational
results of parameters that are believed to have a potentially
continuous range (like the charge and mass of the proton).
(More precisely, parameters like the charge and mass of the proton
are not now known to be limited to discrete values,
so for the sake of argument I shall assume that they can in principle
have continuous ranges,
even though a future theory could conceivably show that they
really are limited to certain discrete values.)

	Then, to one-loop order and ignoring additive numbers like
$1/4\pi$, the results of Marciano and Senjanovi\'c \cite{MarCen}
imply that the $SU(3)\times SU(2)\times U(1)$ inverse
coupling constants, and the inverse electromagnetic fine-structure
constant, run with the energy scale $\mu$,
in the range $m_W < \mu < m_S$ between the weak energy scale $m_W$
and the unification mass scale $m_S$, in the following way,
when $n_g = 3$ and $N_H = 2$:
 \begin{eqnarray}
 \alpha_1^{-1}(\mu) \approx \alpha_1^{-1}(m_S)
    +{2n_g+0.3N_H \over 2\pi}\ln{m_S\over\mu}
    &=& \alpha_1^{-1}(m_S) + {3.3\over\pi}\ln{m_S\over\mu},
 \label{eq:11} \\
 \alpha_2^{-1}(\mu) \approx \alpha_2^{-1}(m_S)
    +{2n_g+0.5N_H-6 \over 2\pi}\ln{m_S\over\mu}
    &\approx & \alpha_1^{-1}(m_S) + {0.5\over\pi}\ln{m_S\over\mu},
 \label{eq:12} \\
 \alpha_3^{-1}(\mu) \approx \alpha_3^{-1}(m_S)
    +{2n_g-9 \over 2\pi}\ln{m_S\over\mu}
    &\approx & \alpha_1^{-1}(m_S) - {1.5\over\pi}\ln{m_S\over\mu},
 \label{eq:13} \\
 \alpha^{-1}(\mu) = {5\over 3}\alpha_1^{-1}(\mu)
    + \alpha_2^{-1}(\mu)
   &\approx &{8\over 3}\alpha_1^{-1}(m_S)+{6\over\pi}\ln{m_S\over\mu}.
 \label{eq:13b}
 \end{eqnarray}
 
	Now the proton mass is set roughly by the scale $\mu$ at which
the $SU(3)$ coupling constant $\alpha_3(\mu)$ becomes large, say
 \begin{equation}
 0 \sim \alpha_3^{-1}(m_p)
   \approx \alpha_1^{-1}(m_S) - {1.5\over\pi}\ln{m_S\over m_p}.
 \label{eq:14}
 \end{equation}
For this equation to be approximately valid despite the fact
that the proton mass $m_p$ is not in the range
$m_W < \mu < m_S$ where Eqs. (\ref{eq:11})-(\ref{eq:13b}) are valid,
I shall assume that on a logarithmic scale,
$m_p$ is fairly close to $m_W$, in comparison with
a much greater logarithmic range where
Eqs. (\ref{eq:11})-(\ref{eq:13b}) are valid,
as is indeed the case in the subuniverse that we observe.

	If we approximate the electromagnetic coupling constant
at zero momentum transfer,
which is the quantity we have been calling $\alpha$,
with the electromagnetic coupling constant at energy scale
$\mu = m_p$, we get from Eqs. (\ref{eq:13b}) and (\ref{eq:14}) that
 \begin{equation}
 \alpha^{-1} \approx \alpha^{-1}(m_p)
             \approx {10\over\pi}\ln{m_S\over m_p}.
 \label{eq:15}
 \end{equation}
 
 	The next approximation we shall make is that on a logarithmic
scale, the unification mass $m_S$ is nearly the Planck mass $m_P$,
so we can replace $\ln{(m_S/m_p)}$ with
$\ln{(m_P/m_p)} = 0.5\ln{\alpha_G^{-1}}$ in Eq. (\ref{eq:15}).
This then gives us our new relation between
the electromagnetic and gravitational fine structure constants
(or between the charge and mass of the proton in Planck units,
as given in the Abstract):
 \begin{equation}
 \alpha^{-1} \approx {5\over\pi}\ln{\alpha_G^{-1}}.
 \label{eq:16}
 \end{equation}
 
 	One can note that if we stick with $n_g = 3$ generations
of quarks and leptons but allow the number
of relatively light Higgs doublets, $N_H$, to be different from
its minimal value of 2, then we get instead
 \begin{equation}
 \alpha^{-1} \approx {9+0.5N_H\over 2\pi}\ln{\alpha_G^{-1}}.
 \label{eq:17}
 \end{equation}

	Alternatively, one can replace the approximate
Eqs. (\ref{eq:16})-(\ref{eq:17}) with the exact equation
 \begin{equation}
 \alpha^{-1} = {N\over 2\pi}\ln{\alpha_G^{-1}}
 \label{eq:18}
 \end{equation}
(taking $N$ to be defined by this equation)
and the approximate relation
 \begin{equation}
 N \approx 9 + 0.5N_H.
 \label{eq:19}
 \end{equation}
Since $N_H$ must be an even integer, the right hand side
of Eq. (\ref{eq:19}) must be an integer, 10 if $N_H = 2$,
whereas we can take the left hand side, $N$,
to be defined by Eq. (\ref{eq:18}).
In our subuniverse, we observe $N \approx 9.7816$ \cite{PDG}.
This is close enough to 10 that we might conclude
that it is likely that our assumption of $N_H = 2$ is correct
for our subuniverse.

	Incidentally, to get some feel for the accuracy
of the approximations above, one may note that at the same
level of approximation and truncation of the MSSM
renormalization group equations \cite{MarCen},
one gets that the weak-mixing angle $\theta_W$ gives
 \begin{equation}
 \sin^2\theta_W \approx 0.2,
 \label{eq:20}
 \end{equation}
about 13\% less than the observed value 0.231 in our subuniverse
\cite{PDG}.

\section{Combining the two relations between $\alpha$ and $\alpha_G$}

	Now we can combine some form of the Carter-Carr-Rees
anthropic relation between $\alpha$ and $\alpha_G$
with the new MSSM renormalization group approximate
Eq. (\ref{eq:16}) to derive approximate estimates for values of both
$\alpha$ and $\alpha_G$ that a typical observer might be expected
to see.

	If we use the basic Carter-Carr-Rees relation above,
Eq. (\ref{eq:7}), $\alpha_G \sim \alpha^{20}$,
then we get for the electromagnetic fine structure constant $\alpha$
or the charge $e = \sqrt{\alpha}$ of the proton in Planck units,
 \begin{equation}
 \alpha\ln{\alpha} = 2 e^2\ln{e} \sim -{\pi\over 100},
 \label{eq:21}
 \end{equation}
with the solution
 \begin{equation}
 \alpha^{-1} \sim 162 \, \approx \, 1.18\,\alpha_{\mathrm{exp}}^{-1}\,.
 \label{eq:22}
 \end{equation}
The prediction for a typical observed value of the charge
of the proton is then
 \begin{equation}
 e \sim 0.0786 \approx 1.47 \times 10^{-19} \, \mathrm{coulomb}
   \approx 0.920 \, e_{\mathrm{exp}},
 \label{eq:24}
 \end{equation}
within about 8\% of the experimental value
observed in our subuniverse near our present time and location,
$e_{\mathrm{exp}} = \sqrt{\alpha_{\mathrm{exp}}} \approx 0.085424543
   \approx 1.6021765 \times 10^{-19}$ coulomb \cite{PDG}.
   
	Similarly, we can combine Eqs. (\ref{eq:7}) and (\ref{eq:16})
to get an equation for the gravitational fine structure constant
$\alpha_G$ or the mass $m_p = \sqrt{\alpha_G}$ of the proton
in Planck units,
 \begin{equation}
 \alpha_G^{1/20}\ln{\alpha_G} = 2 m_p^{1/10}\ln{m_p}
                     \sim -{\pi\over 5}.
 \label{eq:25}
 \end{equation}
This has the solution
 \begin{equation}
 \alpha_G^{-1} \sim 1.54 \times 10^{45} \sim 2^{147}
 \sim 907\,000 \, \alpha_{G\ \mathrm{exp}}^{-1},
 \label{eq:26}
 \end{equation}
or
 \begin{equation}
 m_p = \sqrt{\alpha_G} \sim 8.1 \times 10^{-23}
 \approx 1.76 \times 10^{-30} \, \mathrm{kg}
 \approx 0.00105 \, m_{p\ \mathrm{exp}}\,.
 \label{eq:27}
 \end{equation}

	In this case the agreement is not so good,
as the estimated mass of the proton
comes out to be only about 0.1\% of the value
we observe in our subuniverse.
The large magnitude of the error can be attributed
to the the large exponent of 20 in
the basic Carter-Carr-Rees relation of Eq. (\ref{eq:7}),
$\alpha_G \sim \alpha^{20}$.
This makes it so that a relatively small error
in the estimate for $\alpha$ can get converted
into a relatively large error in $\alpha_G$.
However, since $m_p$ is so small in Planck units,
it might be more natural to make the comparison
on a logarithmic scale, in which case one gets
 \begin{equation}
 {\ln{m_p}\over \ln{m_{p\ \mathrm{exp}}}} \sim 1.156,
 \label{eq:28}
 \end{equation}
which has an error of less than 16\%.

	Although of course it is a cheat
for getting a true estimate,
one can improve the estimate by using
the fudged Carter-Carr-Rees relation, Eq. (\ref{eq:8}),
$\alpha_G \sim 10^4\,\alpha^{20}$,
along with the MSSM renormalization group approximate
Eq. (\ref{eq:16}) to get the fudged improved equations
 \begin{equation}
 \alpha\ln{(10^{0.2}\,\alpha)} = 2 e^2\ln{(10^{0.1}\,e)}
  \sim -\,0.01\,\pi,
 \label{eq:29}
 \end{equation}
 \begin{equation}
 \alpha_G^{0.05}\ln{\alpha_G} = 2 m_p^{0.1}\ln{m_p}
                     \sim -10^{0.2}\,0.2\,\pi.
 \label{eq:30}
 \end{equation}
The approximate numerical solutions of these fudged equations are
 \begin{equation}
 \alpha^{-1} \sim 143.4 \approx 1.046\,\alpha_{\mathrm{exp}}^{-1}\,,
 \label{eq:31}
 \end{equation}
 \begin{equation}
 e \sim 0.08351 \approx 1.566 \times 10^{-19} \, \mathrm{coulomb}
   \approx 0.9775 \, e_{\mathrm{exp}}\,,
 \label{eq:32}
 \end{equation}
 \begin{equation}
 \alpha_G^{-1} \sim 1.35 \times 10^{39} \sim 2^{130}
 \sim 7.99 \, \alpha_{G\ \mathrm{exp}}^{-1}\,,
 \label{eq:33}
 \end{equation}
 \begin{equation}
 m_p = \sqrt{\alpha_G} \sim 2.72 \times 10^{-20}
 \approx 5.92 \times 10^{-28} \, \mathrm{kg}
 \approx 0.354 \, m_{p\ \mathrm{exp}}\,.
 \label{eq:34}
 \end{equation}

	Thus the fudged equations give an estimated charge
of the proton within about 2.25\% of the observed value
nearby in our subuniverse, and an estimated mass of the proton
within a factor of 3 of the observed value.
Of course, the factor of 10 in Eq. (\ref{eq:4}),
which gets raised to the 4th power in Eq. (\ref{eq:8}),
was just an order-of-magnitude estimate that was no doubt biased
by the observed value of this factor in our subuniverse,
so it is not expected to be reliable.
However, it does show that if one combines my
new MSSM renormalization group approximate Eq. (\ref{eq:16})
(having no continuous free parameters)
with the previously-published anthropic equations with
this previously-published fudge factor,
one gets a remarkable agreement with the observed
charge of the proton, nearly 98\% of the observed value,
and reasonably good agreement with the observed mass,
nearly 98\% of its logarithm.

	More importantly, since it shows that one needs
a fudge factor, like the factor of 10 in Eq. (\ref{eq:4}),
that is of the same general
order of magnitude as unity to get good agreement with
the observations in our subuniverse, it does make it
plausible to conjecture that when we gain sufficient
knowledge to be able to calculate more precisely
the statistical distribution of observed values
of the charge and mass of the proton over all subuniverses,
the values we observe in our subuniverse will turn out
to be typical, even if the values can vary
widely over the full range of subuniverses.

	If one cheats even more and uses the
MSSM renormalization group approximate Eq. (\ref{eq:16})
with the 2-3-4 formula (\ref{eq:9b}),
one gets an even better fit to our observations,
with the charge of the proton being predicted
to be 0.986 of the observed value,
and the mass of the proton being predicted to be
0.760 of the observed value.
(This replacement is equivalent to replacing
the Carr-Rees fudge factor of 10,
in Eq. (\ref{eq:8}) where it is raised to the 4th power,
with $2^{15/4} \approx 13.4543$.)
Since the new fudge factor of $2^{3/4}$ in the 2-3-4 formula (\ref{eq:9b})
was chosen as a simple expression close to the numerical
value for that equation to be true in our subuniverse,
it cannot be counted as a prediction at all.
Then the main error in the `predicted' charge and mass of the
proton comes from the
MSSM renormalization group approximate Eq. (\ref{eq:16}).

	Alternatively, one can choose $F$ in Eq. (\ref{eq:9})
so that when it is combined with Eq. (\ref{eq:16})
to get separate equations for $\alpha$ and $\alpha_G$,
 \begin{equation}
 \alpha\ln{(F\alpha)} = -\,0.01\pi,
 \label{eq:36}
 \end{equation}
 \begin{equation}
 \alpha_G^{1/20}\ln{\alpha_G} = -\,0.2\,\pi F,
 \label{eq:37}
 \end{equation}
one fits the observed value of either $\alpha$ or $\alpha_G$
and then predicts the other one.

	For example, if one uses our observed value of $\alpha^{-1}$,
$\alpha_{\mathrm{exp}}^{-1} \approx 137.036000$ \cite{PDG},
and solves Eq. (\ref{eq:36}) for
 \begin{equation}
 F = \alpha_{\mathrm{exp}}^{-1}\exp{(-\,0.01\pi\alpha_{\mathrm{exp}}^{-1})}
   \approx 1.8498985
 \label{eq:38}
 \end{equation}
and then inserts this into Eq. (\ref{eq:37})
to solve for $\alpha_G = m_p^2$,
one gets a value for the mass of the proton that
is too large by a factor of about 2.615,
rather oppositely to the previous predicted values
that were all too small.

	On the other hand, if one uses our observed value of $\alpha_G$,
$\alpha_{G\ \mathrm{exp}} \approx 2^{-126.993}$ \cite{PDG},
and solves Eq. (\ref{eq:37}) for
 \begin{equation}
 F = -{5\over\pi}\,\alpha_{G\ \mathrm{exp}}^{0.05}
      \ln{\alpha_{G\ \mathrm{exp}}}
   \, \approx 1.7179
 \label{eq:39}
 \end{equation}
and then inserts this into Eq. (\ref{eq:36})
to solve for $\alpha = e^2$,
one gets a value for the charge of the proton
that is 0.989 of the observed value.

	A possibility very similar to this last one is to
replace the 10 in Eqs. (\ref{eq:29})-(\ref{eq:30}) with 15, which is
equivalent to using $F = 15^{0.2} \approx 1.7188$ in
Eqs. (\ref{eq:36})-(\ref{eq:37}).  Then
one gets $e \approx 0.9891 \, e_{\mathrm{exp}}$
and $m_p \approx 1.0069 \, m_{p\ \mathrm{exp}}$,
so then the errors are less than 1.1\% and 0.7\% respectively.

	To fit our observed values
of both the charge and mass of the proton to better than 99\% accuracy,
we need not only a single fudge factor like $F$ in Eq. (\ref{eq:9}),
but also a second fudge factor in Eq. (\ref{eq:16}),
such as the $N$ in Eq. (\ref{eq:18}).
As noted previously, if we set $N \approx 9.7816$,
as well as $F \approx 1.6803$, then we can reproduce
our observed values of both the charge and mass of the proton.
However, with 3 generations of quarks and leptons,
and the minimal number (2) of relatively light Higgs doublets,
the coefficient $N$ in Eq. (\ref{eq:18}) is predicted to be precisely 10.
Thus if we did put a fudge factor into that equation,
it should be somewhere else.

	The main error of Eq. (\ref{eq:16}) seems to come from
the approximations that on a logarithmic scale,
the unification mass is close to the Planck mass
and the weak energy scale is close to the proton mass.
The first of these errors tends to increase the right hand side
above the left hand side, and the second tends to decrease it.
If we wildly conjecture that the ratios of the unification mass
to the Planck mass and of the weak energy scale to the proton mass
might be expected anthropically to go as some powers of the running
coupling constants at those scales,
then we might na\"{\i}vely expect that a first correction to
Eq. (\ref{eq:16}) would be to replace it with
 \begin{equation}
 \alpha^{-1} \approx {5\over\pi}\ln{(\alpha^p\alpha_G^{-1})}
 \label{eq:40}
 \end{equation}
for some exponent $p$ that might conceivably be predictable.
If we then set $F = e^q$ for another exponent $q$
(where here and in the following equation $e$ means the base
of the natural logarithms rather than the charge of the proton),
then Eq. (\ref{eq:9}) becomes
 \begin{equation}
 \alpha_G = (e^q \alpha)^{20}.
 \label{eq:41}
 \end{equation}
Here the factor of $5/\pi$ in Eq. (\ref{eq:40})
and the exponent of 20 in Eq. (\ref{eq:41})
are determined by discrete parameters of the crude theory
used to derive these equations,
but $p$ and $q$ are not so determined, at least not yet.
The simplest version of the argument above would
just suggest that $p$ and $q$ should both be roughly 0.

	Using both fudge factors $p$ and $q$
in Eqs. (\ref{eq:40}) and Eq. (\ref{eq:41}),
we can fit the observations in our subuniverse
with $p \approx 0.390782$ and $q \approx 0.518993$.

	Now the point of a complete anthropic argument
would not be to predict precise values for $p$ and $q$
(which would give precise values for the charge and mass of the proton),
but rather to predict a joint statistical distribution for their
observed values over the entire universe.
The simple argument given above just suggests that typical observers,
if they are like us in living on planets, having complex chemistry, etc.,
might expect to see values of $p$ and $q$ that are not
too different from zero
(i.e., having magnitudes not very large in comparison with unity).
This does fit the observations in our subuniverse,
where both $p$ and $q$ are even somewhat smaller than unity.
In this way our observations are consistent with
the simple argument given above.

	Of course, this consistency just means that the simple argument
passed its first test in not being falsified
where it conceivably could have been,
but the consistency does not confirm the basic truth
of the simple argument.
For example, we do not yet know whether the charge and mass of the proton
really do have a distribution of different values over all subuniverses.
However, it is at least encouraging that one can find a simple argument
of this form that does so well in giving rough values for our
observations of the charge and mass of the proton.

	I have benefited from discussions on this topic with Bruce
Campbell, Gerald Cleaver, Valeri Frolov, Gordon Kane, Sharon Morsink,
Roger Penrose, Lenny Susskind, and no doubt others.  This research was
supported in part by the Natural Sciences and Engineering Research Council
of Canada.


\end{document}